%% file: CPM_1bit.tex
\documentclass[12pt,onecolumn]{IEEEtran}

\usepackage[nocompress]{cite}
\usepackage{epsfig,xcolor}
\usepackage{amsmath,graphicx,subfig}
\usepackage{multicol}
\usepackage{multirow}
\usepackage{balance}
\usepackage{nicefrac}
\usepackage{adjustbox}
\usepackage{enumitem}
\usepackage{url}

\usepackage{mathtools,amssymb,cuted}

\usepackage{algorithm,algorithmic}

\newcommand{\cev}[1]{\reflectbox{\ensuremath{\vec{\reflectbox{\ensuremath{#1}}}}}}
\def\BibTeX{{\rm B\kern-.05em{\sc i\kern-.025em b}\kern-.08em
    T\kern-.1667em\lower.7ex\hbox{E}\kern-.125emX}}

\begin{document}

\title{\huge Study of Symbol Error Probability Constrained Precoding with Zero-Crossing Modulation for Wireless Systems with 1-Bit ADCs}

\author{Diana~M.~V.~Melo, Lukas~T.~N.~Landau,				 Rodrigo~C.~de~Lamare}

\maketitle

\begin{abstract}
The next generation of wireless communications systems will employ new frequency bands such as those in the upper midband, millimeter-wave and sub-terahertz frequency bands. The high energy consumption of analog-to-digital converters resulting from their high resolution constituted a major limitation for future wireless communications systems, which will require low energy consumption and low-complexity devices at the transmitter and at the receiver.  In this regard, we present a novel precoding method based on quality of service constraints for a multiuser multiple-input multiple-output downlink system with 1-bit quantization and oversampling. For this scenario, we consider the time-instance zero-crossing modulation, which conveys the information into the zero-crossings of the signals. Unlike prior works the proposed constraint is given in terms of the symbol error probability related to the minimum distance to the decision threshold and is included in the proposed optimization problem that is used in the design of the precoder. Simulation results illustrate the performance of the proposed precoding method evaluated under different parameters and scenarios.

 \end{abstract}

\begin{IEEEkeywords}
Precoding, energy-efficient communications, zero-crossing modulation.
\end{IEEEkeywords}

\IEEEpeerreviewmaketitle

\input{Introduction}

\input{System_Model}
\input{TIZX}

\input{QOS}
\input{semianalitical}

\input{numerical_Results}
\input{Conclusions}

\bibliographystyle{IEEEtran}
\bibliography{bib-refs}

\end{document}

%% file: Introduction.tex
\section{Introduction}
The next generation of wireless communications systems will employ new frequency bands such as those in the upper midband, millimeter-wave and sub-terahertz frequency bands and support a massive number of devices \cite{Rappaport2019} as in Internet of Things (IoT) scenarios \cite{Gupta_2015} and reach higher data rates \cite{Viswanathan_2020}. The transmission of very high data rates corresponds to a big challenge in terms of the design of energy-efficient analog
to-digital converters (ADCs) because their energy consumption increases exponentially with amplitude resolution \cite{Murmann_2009} and quadratically with the sampling rate for bandwidths that exceed 300MHz \cite{Murmann_ADC}.
A well-known and widely adopted strategy to decrease the power consumption of each ADC is to low-resolution quantization and even consider 1-bit quantization \cite{bbprec,1bitidd,dqalms,dqarls,dynovs,comp}. In addition to the reduction of energy consumption, this strategy can reduce the complexity of transceivers and their devices as certain tasks like automatic gain control can potentially be discarded.

The loss of information carried in the amplitudes can be compensated by increasing the sampling rate \cite{Gilbert_1993} of the signal to be processed. In a noise-free case, it has been shown that rates of $\log_2(M_\mathrm{Rx}+1)$ bits per Nyquist interval are achievable by $M_\mathrm{Rx}$-fold oversampling with respect to (w.r.t.) the Nyquist rate \cite{Shamai2_1994}.
To this end, the information must be encoded into the temporal samples with one-bit resolution. Methods based on 1-bit quantization and oversampling have been introduced in \cite{Landau_2014, Son_2019,1bitcpm,dynovs}. In the work of \cite{Shamai2_1994} the constructed bandlimited transmit signal conveys the information into zero-crossing patterns. In this context, modulation schemes based on zero-crossing have been proposed in \cite{Peter_2021, Peter_2020, Bender_2019} with run-length limited (RLL) transmit sequences and in the work of \cite{Viveros_2023} with the time instance zero-crossing (TI ZX) modulation. The TI ZX modulation from \cite{Viveros_2023} encodes the information into the time-instance of zero-crossings in order to reduce the number of zero-crossings of the signal. In the context of multiple-antenna systems, important precoding approaches for multi-input multi-output (MIMO) channels have been studied for systems with 1-bit quantization modulation based on the maximization of the minimum distance to the decision threshold (MMDDT) \cite{Viveros_2023}, minimum mean squared error (MMSE) \cite{Viveros_2023, Amine_2016, Jacobson_2016}, state-machine based waveform design optimization \cite{Viveros_2024} and quality of service (QOS) constraint \cite{Viveros_ssp_2021}. The proposed method in \cite{Viveros_ssp_2021}  minimizes the transmit power while taking into account quality of service constraints in terms of the minimum distance to the decision threshold. Some of these methods improve performance when combined with faster-than-Nyquist (FTN) signaling \cite{Mazo_1975}. Moreover, in the approach presented in \cite{Viveros_2023} a spectral efficiency lower bound is presented for the TI ZX modulation. Furthermore, in the work in \cite{Viveros_2023} an analytical method was introduced for the TI ZX modulation with MMSE precoding. In addition, the study in \cite{Erico_2023} proposed a branch-and-bound method with QOS for a solution that attains a target symbol-error probability. 

 In this work, we present a bandlimited multiuser multiple-antenna downlink wireless system with 1-bit quantization and oversampling, which employs the TI ZX modulation concept \cite{Viveros_2023}. In particular, a semi-analytical symbol error rate upper bound is incorporated in the QOS constraint method with an established minimum distance to the decision threshold \cite{Viveros_ssp_2021}. Unlike our previous work in \cite{Viveros_H_2021}, the proposed precoding method is formulated such that the constraint is given in terms of a target SER. Moreover, we adopt the Gray coding for the TI ZX modulation adopted in the work of \cite{Viveros_2023}, an approximately BER can also be defined as constraint.
 
The rest of this paper is structured as follows: Section~\ref{sec:system_model} details the adopted system model. Afterwards, in Section~\ref{sec:TIZX} we present the TI ZX modulation and the detection stage of the system under consideration. Section~\ref{sec:QOS} presents the optimization problem for QOS temporal precoding method. Then, the performance bound is introduced in Section~\ref{sec:bound}. Simulation results are presented in Section~\ref{sec:num_results}, whereas Section~\ref{sec:conclusiones} draws the concluding remarks of this work.

Notation: In the paper all scalar values, vectors and matrices are represented by $a$, ${\boldsymbol{x}}$ and ${\boldsymbol{X}}$, respectively.

%% file: System_Model.tex
\section{Multiuser Multiple-Antenna System Model}
\label{sec:system_model}

\begin{figure}[t]
\begin{center}
\includegraphics[height=8cm, width=17cm]{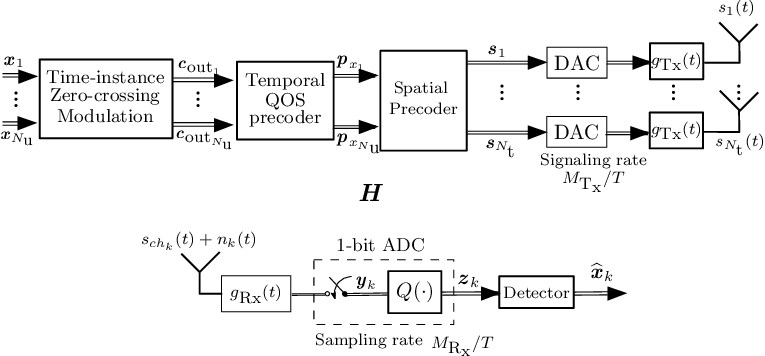}
\caption{Multiuser MIMO system model}
\label{fig:system_mimomodel}       
\end{center}
\end{figure}

The multiuser multiple-antenna system model is considered in the downlink with $N_{\text{t}}$ transmit antennas and $N_{\text{u}}$ single antenna users as depicted in Fig.~\ref{fig:system_mimomodel}. The vector $\boldsymbol{x}_{k} \in \mathbb{C}$ for user $k$, with $N$ symbols corresponds to the transmit vector that is mapped into the TI ZX modulated pattern $\boldsymbol{c}_{\textrm{out}} \in \mathbb{C}^{N_{\textrm{tot}}\times 1}$ where $N_{\textrm{tot}} = N M_{\textrm{Rx}}+1$ and $M_{\textrm{Rx}}$ corresponds to the oversampling factor. Then, the space-time precoding vector $\boldsymbol{p}_{x} \in \mathbb{C}^{N_{\textrm{q}}{N_{\textrm{T}}}\times 1}$ is obtained,  where $N_{\text{q}} = M_{\text{Tx}}N+1$.

The signaling rate factor $M_{\text{Tx}} >1$ corresponds to faster-than-Nyquist signaling and is related to the oversampling factor by $MM_{\text{Tx}} = M_{\text{Rx}}$. At the BS ideal digital-to-analog converters (DACs) are considered.  The transmit pulse shaping filter $g_{\text{Tx}}(t)$ in discrete time, with unit  energy normalization $a_{\textrm{Tx}}=  (T /  M_{\text{Tx}} )^{1/2} $, is represented by the Toeplitz matrix  $\boldsymbol{G}_{\textrm{Tx}}$  given by
\begin{align}
\label{eq:GTx}
\boldsymbol{G}_{\textrm{Tx}}=  a_{\textrm{Tx}} \begin{bmatrix}
\left[ \ \boldsymbol{g}_{\text{Tx}}^T \ \right]  \ 0 \cdots \ \ \ 0  \\
0 \ \left[ \ \boldsymbol{g}_{\text{Tx}}^T \ \right] \ 0 \cdots 0 \\
\ddots  \ddots \ddots    \\
0 \cdots \ \ \  0 \ \left[ \ \boldsymbol{g}_{\text{Tx}}^T  \ \right]   
\end{bmatrix}_{N_{\text{tot}}  \times 3N_{\text{tot}} }
\text{.}
\end{align} 

 At the receiver, a pulse shaping filter and 1-bit analog-to-digital converter, process the signal of the $k$th user. The receive filter $g_{\text{Rx}}(t)$  is represented by the Toeplitz matrix 
\begin{align}
\label{eq:GRx}
\boldsymbol{G}_{\textrm{Rx}}=  a_{\textrm{Rx}} \begin{bmatrix}
\left[ \ \boldsymbol{g}_{\text{Rx}}^T \ \right]  \ 0 \cdots \ \ \ 0  \\
0 \ \left[ \ \boldsymbol{g}_{\text{Rx}}^T \ \right] \ 0 \cdots 0 \\
\ddots  \ddots \ddots    \\
0 \cdots \ \ \  0 \ \left[ \ \boldsymbol{g}_{\text{Rx}}^T  \ \right]   
\end{bmatrix}_{N_{\text{tot}}  \times 3N_{\text{tot}} }
\text{,}
\end{align} 
with
$\boldsymbol{g}_{\text{Rx}} = [ g_{\text{Rx}} (-T  ( N + \frac{1}{M_{\text{Rx}}}  )  ),
g_{\text{Rx}}(-T  ( N + \frac{1}{M_{\text{Rx}}} )  + \frac{T}{M_{\text{Rx}}}  ), \ldots,$
$ g_{\text{Rx}} (T ( N + \frac{1}{M_{\text{Rx}}} )  ) ]^T $ as the coefficients of the ${g}_{\text{Rx}} (t)$ filter and $a_{\textrm{Rx}}=  (T /  M_{\text{Rx}} )^{1/2} $  that corresponds to  unit  energy normalization.
The channel matrix $\boldsymbol{H} \in \mathbb{C}^{N_{\textrm{u}}\times N_{\textrm{t}}}$ describes a frequency flat fading channel. 

The effects of pulse shaping filtering are given by the combined waveform determined by the transmit and receive filters which is described by  $ v (t) = g_{\text{Tx}}(t) * g_{\text{Rx}}(t) $. The combined waveform is represented by the Toeplitz matrix $\mathbb{\boldsymbol{V}}$,
 \begin{align}
\label{eq:MatrixV_offset}
  \mathbb{\boldsymbol{V}} = \;
   \begin{bmatrix}
      v\left ( 0 \right ) & v\left ( \frac{T}{M_{\text{Rx}}} \right ) & \cdots &  v\left (T N  \right ) \\
      v\left ( -\frac{T }{M_{\text{Rx}}}  \right ) & v\left ( 0 \right ) & \cdots &  v\left (T \left ( N-\frac{1}{M_{\text{Rx}}} \right ) \right ) \\
      \vdots & \vdots & \ddots & \vdots \\
			v\left (-T N \right ) &  v\left (  T \left ( -N+\frac{1}{M_{\text{Rx}}} \right )  \right ) & \cdots &  v\left (0 \right )
   \end{bmatrix}_{N_{\text{tot}} \times N_{\text{tot}}}
\end{align}
The matrix $\boldsymbol{U} \in \mathbb{R}^{N_{\text{tot}} \times N_{\text{q}}}$, describes the $M$-fold upsampling operation which relates different signaling
and sampling rates
\begin{align}
\label{eq:Matrizu}
\boldsymbol{U}_{m,n}=
\begin{cases}
  1,  & \textrm{for} \quad m = M \cdot \left ( n-1 \right )+1\\
  0, &  {\textrm{else.}}
\end{cases}
\end{align}
The received signal is quantized and vectorized such that $\boldsymbol{z} \in \mathbb{C}^{N_{\text{u}}N_{\text{tot}}}$ is obtained as
\begin{equation}
\boldsymbol{z} = Q_{1}\left ( \boldsymbol{y} \right ),
\end{equation}
where
\begin{equation}
\boldsymbol{y} = \left (\boldsymbol{H}\boldsymbol{P}_{\text{sp}} \otimes \boldsymbol{I}_{N_{\text{tot}}}\right )\left (\boldsymbol{I}_{\text{N}_{\text{u}}} \otimes \boldsymbol{V}\boldsymbol{U}\right )\boldsymbol{p}_{\text{x}}  + \left (\boldsymbol{I}_{\text{N}_{\text{u}}} \otimes \boldsymbol{G}_{\text{Rx}}\right )\boldsymbol{n}\text{.}
\end{equation}

The stacked vector $\boldsymbol{p}_{\text{x}}= \left [\boldsymbol{p}_{\text{x}_1}^{T},\boldsymbol{p}_{\text{x}_2}^{T},\cdots,\boldsymbol{p}_{\text{x}_k}^{T},\cdots,\boldsymbol{p}_{\text{x}_{N_{\text{u}}}}^{T} \right ]^{T}$ with $\boldsymbol{p}_{\text{x}_k} \in \mathbb{C}^{N_{\text{q}}}$ corresponds to the  temporal precoding vector. The vector  $\boldsymbol{n} \in \mathbb{C}^{N_{\text{u}}3N_{\text{tot}}}$ represents the complex Gaussian noise vector with zero mean and variance $\sigma_{n}^{2}$. 
Considering perfect channel state information, the conventional spatial Zero Forcing (ZF) precoding matrix \cite{SpencerHaardt_2004} is defined as
\begin{align}
\boldsymbol{P}_{\text{sp}} = c_{\text{zf}}\boldsymbol{P}_{\text{zf}}
\end{align}
where
\begin{align}
\boldsymbol{P}_{\text{zf}} = \boldsymbol{H}^{H}\left ( \boldsymbol{H}\boldsymbol{H}^{H}\right )^{-1}
\end{align}
The scaling factor $c_{\text{zf}}$ is given by
\begin{align}
c_{\text{zf}} =  \sqrt{\left (  N_{\text{u}}/ \mathrm{trace} \left (  \left ( {\boldsymbol{H}}{\boldsymbol{H}}^{H} \right )^{-1} \right )  \right )} \text{.}
\end{align}
Several alternative precoding structures \cite{siprec,bbprec,gbd,wlbd,cqabd,rsbd,mbthp,rsthp} could be considered in this system.
The next section briefly describes the time-instance zero-crossing modulation \cite{Viveros_2023}.
The next section briefly describes the time-instance zero-crossing modulation \cite{Viveros_2023}.
 

%% file: TIZX.tex
\section{Modulation Based on Time Instances of Zero Crossings}
\label{sec:TIZX}

This section details the Time Instances of Zero Crossings (TI ZX) modulation proposed in \cite{Viveros_2023}. The TI ZX modulation conveys the information into the time instances of zero-crossings and considers the absence of zero crossings as a valid mapping from bits to sequences. Taking into account $R = 1+ M_{\text{Rx}}$ unique symbols, each symbol $\boldsymbol{x}_{k,i}$, where $i=1,2,\cdots, N$  is mapped to a  codeword $\boldsymbol{c}_{s_{k,i}}$ of $M_{\text{Rx}}$ binary samples which convey the information according to the time instances in which the zero crossing occurs or not within the symbol interval, as related in the $c_\text{map}$ assignment in Table~\ref{tab:cmap}. The desired output sequence $\boldsymbol{c}_{\text{out}_{k}}$ is obtained by the concatenation of the mapping sequences of each transmit symbol.

\begin{figure}[t]
\begin{center}
\includegraphics[height=4cm, width=11cm]{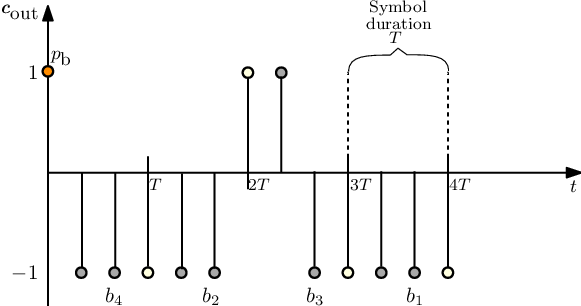}
\caption{Representation of the $\boldsymbol{c}_{\text{out}}$ sequence for $M_{\textrm{Rx}}=3$.}
\label{fig:ejemploCout}       
\vspace{-0.8em}
\end{center}
\vspace{-0.6em}
\end{figure}

The codewords $\boldsymbol{c}_{s_{k, i}}$ of each transmit symbol need to be concatenated to construct the vector $\boldsymbol{c}_{\text{out}_{k}}$, then the last sample of the previous segment must be taken into consideration to generate the appropriate codeword that meets the assignments established in $\boldsymbol{c}_{\text{map}}$. In the same way, a pilot sample needs to be added at the beginning of the sequence to map the first transmit symbol. The last means that each single symbol can be mapped to two different codewords containing the same information based on zero crossings. The Gray coding for TI ZX modulation proposed in \cite{Viveros_2023} is also considered for this study. 
\begin{table}
\caption{{Zero-crossing assignment $\boldsymbol{c}_{\text{map}}$}} 
\label{tab:cmap} 
\normalsize
\begin{center}
\scalebox{0.65}{
\begin{tabular}{|c|l|}
\hline
\multicolumn{2}{|c|}{$\boldsymbol{c}_{\text{map}}$}                                                    \\ \hline
{\color[HTML]{000000} symbol} & \multicolumn{1}{c|}{{Zero-crossing assignment}} \\ \hline
$b_1$                         & No zero-crossing                                                                            \\ \hline
$b_2$                         & Zero-crossing in the $M_{\text{Rx}}$ interval                                                \\ \hline
$b_3$                         & Zero-crossing in the $M_{\text{Rx}}-1$ interval                                              \\ \hline
$\vdots$                      &  \multicolumn{1}{c|}{$\vdots$}                                                                \\ \hline
$b_{R_{\text{in}}-1}$         & Zero-crossing in the second interval                                                         \\ \hline
$b_{R_{\text{in}}}$           & Zero-crossing in the first interval                                                          \\ \hline
\end{tabular}}
\end{center}
\end{table}
The mapping process is done separately and in the same way for the in-phase and quadrature components of the symbols. 
{Fig.~\ref{fig:ejemploCout} shows an example of the construction of the $\boldsymbol{c}_{\text{out}}$ sequence for $M_{\textrm{Rx}}=3$, when the transmitted symbols are $\boldsymbol{x} = \left [ b_4,b_2,b_3,b_1 \right ] $.}
The detection process is performed as the mapping process, separately and similarly for the in-phase and quadrature components of the received sequence of samples $\boldsymbol{z}_{k}$. The process is based on the Hamming distance metric \cite{LarsonB_2017}.
The received sequence $\boldsymbol{z}_k$ is segmented into $N$ subsequences $\boldsymbol{z}_{\textrm{b}_i}= [ \rho_{i-1}, \boldsymbol{z}_i ]^T \in \{+1,-1\}^{M_\mathrm{Rx}+1}$, 
where $\rho_{i-1}$ corresponds to the last sample of the previous  $\boldsymbol{z}_{\textrm{b}_{i-1}}$ subsequence. Then the backward mapping process is defined as $\cev{d}: \boldsymbol{z}_{\textrm{b}_i} \rightarrow [\rho_{i-1}, \boldsymbol{c}_{\textrm{s}i}^T]$ \cite{viveros2020_ICASSP,viveros2020_WSA}. With no distortion, each subsequence $\boldsymbol{z}_{\textrm{b}_i}$ has associated a codeword according to $\boldsymbol{c}_{\text{map}}$, and the detection is performed in a simple way.

With the presence of noise, invalid segments $\boldsymbol{z}_{\textrm{b}_i}$ may be presented. Therefore, the Hamming distance metric is required \cite{LarsonB_2017}  which is defined as
\begin{equation}
    \hat{\boldsymbol{x}}_i = \cev{d}(\boldsymbol{c}), \text{~~with~~} \boldsymbol{c} = \arg \!\!\!\! \min_{\boldsymbol{c}_\mathrm{map} \in \mathcal{M}} \!\! \mathrm{Hamming}(\boldsymbol{z}_{\textrm{b}_i}, \boldsymbol{c}_\mathrm{map}),
\end{equation}
where $\text{Hamming} \; ( \boldsymbol{z}_{\textrm{b}_i},\boldsymbol{c}_{\text{map}}) = \sum_{n=1}^{M_{\text{Rx}}+1}\frac{1}{2}\left | \boldsymbol{z}_{\textrm{b}_{i,n}}- \boldsymbol{c}_{\text{map},n} \right |$ and  $\boldsymbol{c}_\mathrm{map} = [\rho_{i-1},\boldsymbol{c}_{\textrm{s}_i}]^T$, and $\mathcal{M}$ denotes all valid forward mapping codewords. Future developments might consider alternative receive processing algorithms \cite{jidf,rrber,spa,mfsic,mbdf,bfidd} for detection and interference mitigation.

The next section describes the temporal precoding optimization based on quality of service constraints.

%% file: QOS.tex
\section{Design of the QOS temporal precoding}
\label{sec:QOS}
The precoding method proposed in \cite{Viveros_ssp_2021} minimizes transmit energy ${E_{{0}_k}}$  for a given value of $\gamma$  that corresponds to the minimum distance to the decision threshold. In this section, we consider the design of the QoS temporal precoding and formulate the design problem as an optimization problem next.

The transmitted energy per user can be defined by
\begin{align}
\label{eq_energy}
{\boldsymbol{E}}_{0_k} ={\boldsymbol{p}}_{\text{sp}_{k}}^{\text{H}}{\boldsymbol{p}}_{\text{sp}_{k}}  [ (\boldsymbol{W}{\boldsymbol{p}}_{\text{x}_kI})^{\text{T}}(\boldsymbol{W}{\boldsymbol{p}}_{\text{x}_kI})+
(\boldsymbol{W}{\boldsymbol{p}}_{\text{x}_kQ})^{\text{T}}(\boldsymbol{W}{\boldsymbol{p}}_{\text{x}_kQ})] \notag \text{,}
\end{align}
where  {${\boldsymbol{p}}_{\text{sp}_{k}}$} denotes the $k$-th column of ${\boldsymbol{P}}_{\text{sp}}$ and $ \boldsymbol{W} =  \boldsymbol{G}^{T}_{{\text{Tx}}}\boldsymbol{U}$.

The convex optimization problem is solved separately per user and dimension such that the temporal precoding vector ${\boldsymbol{p}}_{\text{x}_k}$ is obtained.
The convex optimization problem  for a given value $\gamma$ can be expressed as
\begin{equation}
\label{eq:convex1}
\begin{aligned}
& \min_{\boldsymbol{p}_{\textrm{x}_{kI/Q}}}
& &  (\boldsymbol{W}{\boldsymbol{p}}_{\text{x}_{kI/Q}})^{\text{T}}(\boldsymbol{W}{\boldsymbol{p}}_{\text{x}_{kI/Q}})\\
& \text{subject to:}
& & \boldsymbol{B}_{k}\boldsymbol{p}_{\text{x}_{kI/Q}} \preceq  -\gamma \boldsymbol{a} \text{,}
\end{aligned}
\end{equation}
where 
\begin{equation}
\label{eq:convexPar}
\begin{aligned}
& \boldsymbol{B}_{\text{k}} &=& -\beta \left ( \boldsymbol{C}_{\text{k}{I/Q}}\boldsymbol{V}\boldsymbol{U} \right)\\
& \boldsymbol{C}_{\text{k}} &=& \textrm{diag}\left (  \boldsymbol{c}_{\text{out}_{\text{k}I/Q}}  \right )\\ 
& \boldsymbol{a}  &=& \boldsymbol{1} \in \mathbb{R}^{M_\mathrm{Rx}N+1} \text{.}
\end{aligned}
\end{equation}
The subscript $I/Q$ denotes that the problem is solved separately for the in-phase and quadrature components of the signal. 
{The constraint in \eqref{eq:convex1} in terms of $\boldsymbol{B}_{\text{k}}$, ensures that the received signal after quantization is equal to $\boldsymbol{c}_{\mathrm{out}_{kI/Q}}$ in a noise-free case  and $\beta$ refers to the real-valued beamforming gain}. The symbol $\preceq$ in \eqref{eq:convex1} constrains each element of the vector $\boldsymbol{B}_{k}\boldsymbol{p}_{\text{x}_{kI/Q}}$ to be less than or equal to $-\gamma$ such that the minimum distance of the samples of the  received signal to the decision threshold is equal to $\gamma$.
Implicitly, the optimization problem shapes the waveform $y(t)$ at the receiver, which is described in the discrete model by $\boldsymbol{H}\boldsymbol{P}_{\text{sp}}\boldsymbol{V} \boldsymbol{U} \boldsymbol{p}_{\textrm{x}_k}$ for the noiseless case.

Considering the spatial ZF precoder, the total transmit energy ${E_{\textrm{Tx}}}$ can be computed as
\begin{align}
\label{Etx}
{E_{\textrm{Tx}}} = \textrm{trace}\left (   \boldsymbol{P}_{\text{sp}}\boldsymbol{R}_{\textrm{x}_{\textrm{Tx}}}\boldsymbol{R}_{\textrm{x}_{\textrm{Tx}}}^{\textrm{H}}\boldsymbol{P}_{\text{sp}}^{\textrm{H}}  \right ),
\end{align}
where  
\begin{align}
\boldsymbol{R}_{\textrm{x}_{\textrm{Tx}}}  = \left [ (\boldsymbol{G}^{\text{T}}_{{\text{Tx}}}\boldsymbol{U}\boldsymbol{p}_{\textrm{x}_{k1}})^{\text{T}};(\boldsymbol{G}^{\text{T}}_{{\text{Tx}}}\boldsymbol{U}\boldsymbol{p}_{\textrm{x}_{k2}})^{\text{T}}; \cdots ; (\boldsymbol{G}^{\text{T}}_{{\text{Tx}}}\boldsymbol{U}\boldsymbol{p}_{\textrm{x}_{kN_{\text{u}}}})^{\text{T}} \right ].
\end{align}
The next section describes the process to obtain the semi-analytical SER.


%% file: semianalitical.tex
\section{Performance Bound for QoS Precoding }
\label{sec:bound}

This section presents the semi-analytical symbol error rate upper bound for the QoS-based precoding method \cite{Viveros_ssp_2021} with the QoS constraint regarding the minimum distance to the decision threshold $\gamma$. 

Considering $M_{\text{Rx}}= 3$, $4$ different symbols $b_1,b_2,b_3,b_4$ can be transmitted. For $M_{\text{Rx}}= 2$, sequences of symbols are considered such that $8$ different symbols $b_1,b_2,b_3,b_4,b_5,b_6,b_7,b_8$ can be transmitted. The SER is defined as
\begin{equation}
    \text{SER} = \text{P}_{\text{error}}.
\end{equation}
Taking into account the probability of correct detection as $\text{P}_{\text{cd}}$, the $\text{P}_{\text{error}}$ probability is defined as  $\text{P}_{\text{error}}= (1-\text{P}_{\text{cd}})$, 
\begin{align}
    \text{SER} &= 1-\text{P}_{\text{cd}}\\
    &=1- {\text{P}\left ( b \right )}\left (  \sum_{i=1}^{m}\text{P}\left ( \hat{x}_{i}= {b}_{i} |x_{i} ={b}_{i} \right )\right ), 
\end{align}
where $m =8$ and $m =4$ for $M_{\text{Rx}}= 2$ and $M_{\text{Rx}}= 3$, respectively. $\text{P}(b) = 1/4$ for $M_{\text{Rx}}= 3$ and $\text{P}(b) = 1/8$ for $M_{\text{Rx}}= 2$, since all input symbols have equal probability. Considering the worse case, that all $N_{\text{tot}}$ samples of the temporal precoding vector $\boldsymbol{p}_{x}$ are equal to a value $\gamma$, where $\gamma$ corresponds to the minimum distance to the decision threshold, the probability of correct detection $\text{P}$, can be lower bounded with
\begin{equation}
    \text{P}'\left ( \hat{x}_{i}= {b}_{i} |x_{i} ={b}_{i} \right ) \leq \text{P}\left ( \hat{x}_{i}= {b}_{i} |x_{i} ={b}_{i} \right ).
\end{equation}
With this, the SER upper bound is defined as
\begin{equation}
    \text{SER}_{\text{ub}} =1- {\text{P}\left ( b \right )}\left (  \sum_{i=1}^{m}\text{P}'\left ( \hat{x}_{i}= {b}_{i} |x_{i} ={b}_{i} \right )\right ), 
\end{equation}
The probability density function of the $m$-dimensional multivariate normal distribution is
\begin{equation}
\label{mvncdf}
f\left ( \boldsymbol{y},\boldsymbol{\mu},\boldsymbol{\Sigma } \right ) = \frac{1}{\sqrt{\left|\Sigma\right |\left ( 2\pi \right )^m}}\text{exp}\left (-\frac{1}{2}(\boldsymbol{y}-\boldsymbol{\mu})\boldsymbol{\Sigma }^{-1}(\boldsymbol{y}-\boldsymbol{\mu})^{T} \right )\text{,}
\end{equation}
where $\boldsymbol{\mu}$ is the mean vector and $\boldsymbol{\Sigma}$ is the covariance matrix defined as $\boldsymbol{\Sigma} = \text{E}\left \{(\boldsymbol{G}_{\textrm{Rx}}\boldsymbol{n})(\boldsymbol{G}_{\textrm{Rx}}\boldsymbol{n})^{T} \right \}$. 

Considering the received vector $\boldsymbol{y}_{i}$ before quantization associated with the input vector $\boldsymbol{x}_{i}$, the correct detection probability is defined as:
\begin{equation}
\text{P}'\left ( \hat{x}_{i}= {b}_{i} |x_{i} ={b}_{i} \right ) =  \int  \cdots \int_{\mathcal{R}}   f\left ( \boldsymbol{y},\boldsymbol{\mu},\boldsymbol{\Sigma } \right ) dy_{1}dy_{2} \cdots dy_{m-1}dy_{m}.
\end{equation}
The integration regions $\mathcal{R}$ and $\boldsymbol{\mu}$ for each symbol $b_{j}$ are presented in Table~\ref{tab:analyMMDDT3} for $M_{\text{Rx}}= 3$ and in Table~\ref{tab:analyMMDDT2} for $M_{\text{Rx}}= 2$.

\begin{table}[H]
\caption{Integration regions $\mathcal{R}$ for each symbol $b_j$ with $M_{\text{Rx}}= 3$.} 
\label{tab:analyMMDDT3}  
\begin{center}
\begin{tabular}{|c|c|c|cc|}
\hline
\multirow{2}{*}{Symbol} & \multirow{2}{*}{$\boldsymbol{\mu}$}                                                 & \multirow{2}{*}{\begin{tabular}[c]{@{}c@{}}Received sequence\\ $\boldsymbol{z}_{i}$ detected as $b_i$\end{tabular}} & \multicolumn{2}{c|}{$\mathcal{R}$}                                                                                     \\ \cline{4-5} 
                        &                                                                        &                                                                                                                     & \multicolumn{1}{c|}{$x_{l}$}                                      & $x_{u}$                                            \\ \hline
\multirow{3}{*}{$b_1$}  & \multirow{3}{*}{$\left [  \gamma, \; \; \gamma, \; \; \gamma, \; \; \gamma \right ]$}    & $\left [ \; \;1, \; \; 1, \; \; 1, \; \; 1 \right ]$                                                                  & \multicolumn{1}{c|}{$\left [\; \; \;0,\; \;\;\;0,\; \; \;\;0,\; \;\;\;0 \right ]$}                   & $\left [  \infty, \infty, \infty, \infty \right ]$ \\ \cline{3-5} 
                        &                                                                        & $\left [ \; \;1, \; \;1, -1, \; \;1 \right ]$                                                                 & \multicolumn{1}{c|}{$\left [\; \; \;0,\; \;\;\;0, -\infty, \; \; \; \;0 \right ]$}          & $\left [ \infty, \infty, \; \; 0, \infty \right ]$      \\ \cline{3-5} 
                        &                                                                        & $\left [  \; \;1, -1, \; \;1, \; \;1 \right ]$                                                                 & \multicolumn{1}{c|}{$\left [\; \;\;0,-\infty, \; \; \; 0, \; \; \; \;0 \right ]$}           & $\left [  \infty, \;\;0, \infty , \infty \right ]$      \\ \hline
\multirow{2}{*}{$b_2$}  & \multirow{2}{*}{$\left [  \gamma, \; \; \gamma, \; \; \gamma,  -\gamma \right ]$}   & $\left [  \; \;1, \; \; 1, \; \; 1, -1 \right ]$                                                                 & \multicolumn{1}{c|}{$\left [\; \; \;0,\; \;\;0,\; \;\;\;0,-\infty \right ]$}             & $\left [  \infty, \infty, \infty, \; \; 0 \right ]$      \\ \cline{3-5} 
                        &                                                                        & $\left [ \; \; 1,  -1, \; \; 1,  -1 \right ]$                                                                & \multicolumn{1}{c|}{$\left [ \; \; \;0,-\infty,\; \; \;0,-\infty \right ]$}       & $\left [  \infty,\;\;0, \infty , \;\;0 \right ]$       
                                                                      \\ \hline
\multirow{1}{*}{$b_3$}  & \multirow{1}{*}{$\left [  \gamma, \; \; \gamma, -\gamma, -\gamma \right ]$}  & $\left [\; \; 1, \; \;1, -1, -1 \right ]$                                                                & \multicolumn{1}{c|}{$\left [\; \; \;0,\; \; \;0,-\infty,- \infty \right ]$}       & $\left [  \infty, \infty, \;\;0, \;\;0 \right ]$           \\ \hline
\multirow{2}{*}{$b_4$}  & \multirow{2}{*}{$\left [\gamma, -\gamma, -\gamma, -\gamma \right ]$} & $\left [\; \;1, -1, -1, -1 \right ]$                                                               & \multicolumn{1}{c|}{$\left [ \; \; \;0,-\infty,-\infty,-\infty \right ]$} & $\left [\infty, \; \;0 \; \; \;0, \; \;0 \right ]$                 \\ \cline{3-5} 
                        &                                                                        & $\left [\; \;1, -1, -1, \; \;1 \right ]$                                                                & \multicolumn{1}{c|}{$\left [\; \; \;0,-\infty,-\infty,\; \;\;\;0 \right ]$}       & $\left [  \infty, \; \;0 \; \; \;0, \infty \right ]$            \\ \hline
\end{tabular}
\end{center}
\end{table}
\begin{table}[H]
\caption{Integration regions $\mathcal{R}$ for each symbol $b_j$ with $M_{\text{Rx}}= 2$.} 
\label{tab:analyMMDDT2}  
\begin{center}
\begin{tabular}{|c|c|c|cc|}
\hline
\multirow{2}{*}{Symbol} & \multirow{2}{*}{$\boldsymbol{\mu}$}                                                 & \multirow{2}{*}{\begin{tabular}[c]{@{}c@{}}Received sequence\\ $\boldsymbol{z}_{i}$ detected as $b_i$\end{tabular}} & \multicolumn{2}{c|}{$\mathcal{R}$}                                                                                     \\ \cline{4-5} 
                        &                                                                        &                                                                                                                     & \multicolumn{1}{c|}{$x_{l}$}                                      & $x_{u}$                                            \\ \hline
\multirow{4}{*}{$b_1$}  & \multirow{4}{*}{$\left [ \gamma, \; \; \gamma, \; \; \gamma, \; \; \gamma, \; \; \gamma \right ]$}    & $\left [ \; \;1,\; \;1, \; \; 1, \; \; 1, \; \; 1 \right ]$                                                                  & \multicolumn{1}{c|}{$\left [\; \; \;0,\; \; \;0,\; \;\;\;0,\; \; \;\;0,\; \;\;\;0 \right ]$}                   & $\left [ \infty, \infty, \infty, \infty, \infty \right ]$ \\ \cline{3-5} 
                        &                                                                        & $\left [ \; \;1,\; \;1, \; \;1, -1, \; \;1 \right ]$                                                                 & \multicolumn{1}{c|}{$\left [\; \; \;0,\; \; \;0,\; \;\;\;0, -\infty, \; \; \; \;0 \right ]$}          & $\left [ \infty, \infty, \infty, \; \; 0, \infty \right ]$      \\ \cline{3-5} 
                        &                                                                        & $\left [ \; \;1,\; \;1, -1, \; \;1, \; \;1 \right ]$                                                                 & \multicolumn{1}{c|}{$\left [\; \;\;0,\; \;\;0, -\infty, \; \;\; \;0, \; \; \; \;0 \right ]$}          & $\left [ \infty, \infty, \; \; 0, \infty,\infty \right ]$      \\ \cline{3-5} 
                        &                                                                        & $\left [  \; \;1, -1, \; \;1, \; \;1, \; \;1 \right ]$                                                                 & \multicolumn{1}{c|}{$\left [\; \; \;0,-\infty, \; \; \;\; 0,\; \; \;\; 0, \; \; \; \;0 \right ]$}           & $\left [  \infty, \;\;0, \infty , \infty, \infty \right ]$      \\ \hline
\multirow{3}{*}{$b_2$}  & \multirow{3}{*}{$\left [ \gamma, \; \; \gamma, \; \; \gamma, \; \; \gamma,  -\gamma \right ]$}   & $\left [  \; \;1,\; \;1, \; \; 1, \; \; 1, -1 \right ]$                                                                 & \multicolumn{1}{c|}{$\left [\; \; \;0,\; \; \;\;\;0,\; \;\;0,\; \;\;\;0,-\infty \right ]$}             & $\left [  \infty, \infty, \infty,\infty, \; \; 0 \right ]$      \\ \cline{3-5} 
&                                                                        & $\left [ \; \;1,\; \;1, -1, \; \;1, -1 \right ]$                                                                 & \multicolumn{1}{c|}{$\left [\; \;\;0,\; \;\;\;0, -\infty, \; \;\; \;0, -\infty \right ]$}          & $\left [ \infty, \infty, \; \; 0, \infty,\infty \right ]$      \\ \cline{3-5} 
                        &                                                                        & $\left [ \; \; 1,  -1, \; \; 1, \; \; 1,  -1 \right ]$                                                                & \multicolumn{1}{c|}{$\left [ \;\; \;0,-\infty,\; \; \;\;0,\; \; \;\;0,-\infty \right ]$}       & $\left [  \infty,\;\;0, \infty,\infty, \;\;0 \right ]$       
                                                                      \\ \hline
\multirow{2}{*}{$b_3$}  & \multirow{1}{*}{$\left [  \gamma, \; \; \gamma, \; \; \gamma, -\gamma, -\gamma \right ]$}  & $\left [\; \; 1, \; \;1,\; \;1, -1, -1 \right ]$                                                                & \multicolumn{1}{c|}{$\left [\; \; \;0,\; \; \;\;0,\; \; \;\;0,-\infty,- \infty \right ]$}       & $\left [  \infty, \infty,\infty, \;\;0, \;\;0 \right ]$  \\ \cline{3-5}          
&                                                                        & $\left [ \; \;1,-1,\; \;1, -1,  -1 \right ]$                                                                 & \multicolumn{1}{c|}{$\left [\; \;\;0, -\infty, \; \;\; \;0, -\infty,-\infty \right ]$}          & $\left [ \infty,  \; \; 0,\infty, \; \; 0,\; \; 0  \right ]$  \\ \hline
\multirow{1}{*}{$b_4$}  & \multirow{1}{*}{$\left [\gamma, \; \; \gamma, -\gamma, -\gamma, -\gamma \right ]$} & $\left [\; \;1,\; \;1, -1, -1, -1 \right ]$                                                               & \multicolumn{1}{c|}{$\left [ \; \; \;0,\; \; \;0,-\infty,-\infty,-\infty \right ]$} & $\left [\infty,\infty, \; \;0 \; \; \;0, \; \;0 \right ]$                        \\ \hline
\multirow{1}{*}{$b_5$}  & \multirow{1}{*}{$\left [\gamma, \; \; \gamma, -\gamma, -\gamma, \; \; \gamma \right ]$} & $\left [\; \;1,\; \;1, -1, -1, \; \;1 \right ]$                                                               & \multicolumn{1}{c|}{$\left [ \; \; \;0,\; \; \;0,-\infty,-\infty,\; \; \;\;0 \right ]$} & $\left [\infty,\infty, \; \;0 \; \; \;0, \infty \right ]$                        \\ \hline 
\multirow{2}{*}{$b_6$}  & \multirow{2}{*}{$\left [\gamma, - \gamma, -\gamma, -\gamma, \; \;\gamma \right ]$} & $\left [\; \;1, -1,-1, -1, \; \;1 \right ]$                                                               & \multicolumn{1}{c|}{$\left [ \; \; \;0,-\infty,-\infty,-\infty,\; \; \;0 \right ]$} & $\left [\infty, \; \;0 \; \; \;0, \; \;0,\;\;\infty \right ]$                 \\ \cline{3-5} 
                        &                                                                        & $\left [\; \;1, -1,\; \;1, -1, \; \;1 \right ]$                                                                & \multicolumn{1}{c|}{$\left [\; \; \;0,-\infty,\; \; \;0,-\infty,\; \;\;\;0 \right ]$}       & $\left [  \infty, \; \;0, \infty, \; \; \;0, \infty \right ]$            \\ \hline  
\multirow{2}{*}{$b_7$}  & \multirow{2}{*}{$\left [\gamma, - \gamma, -\gamma, -\gamma, -\gamma \right ]$} & $\left [\; \;1,-1, -1, -1, -1 \right ]$                                                               & \multicolumn{1}{c|}{$\left [ \; \; \;0,-\infty,-\infty,-\infty,-\infty \right ]$} & $\left [\infty, \; \;0, \; \; \;0, \; \;0 \; \;0 \right ]$                 \\ \cline{3-5} 
                        &                                                                        & $\left [\; \;1, -1,-1, \; \;1 -1 \right ]$                                                                & \multicolumn{1}{c|}{$\left [\; \; \;0,-\infty,\; \; \;0,-\infty,\; \;\;\;0 \right ]$}       & $\left [  \infty, \; \;0,\; \;\infty, \; \; \;0,\infty  \right ]$            \\ \hline
\multirow{1}{*}{$b_8$}  & \multirow{1}{*}{$\left [\gamma, - \gamma, -\gamma, \; \;\gamma, \; \;\gamma \right ]$} & $\left [\; \;1, -1, -1, \; \;1,\; \;1 \right ]$                                                               & \multicolumn{1}{c|}{$\left [ \; \; \;0,-\infty,-\infty,\; \; \;0,\; \; \;0 \right ]$} & $\left [\infty, \; \;0, \; \; \;0, \; \;\infty,\infty \right ]$                          \\ \hline
\end{tabular}
\end{center}
\end{table}
Note that, invalid codewords are also detected as the received symbol $b_{j}$, therefore, also invalid codewords are included in Table~\ref{tab:analyMMDDT3} and Table~\ref{tab:analyMMDDT2}. Moreover, due to symmetry, only the codewords for positive $\rho$ are considered. 

Since a given SER value is related to $\gamma$, the optimization problem in \eqref{eq:convex1} can be reformulated as
\begin{equation}
\label{eq:convex1b}
\begin{aligned}
& \min_{\boldsymbol{p}_{\textrm{x}_{kI/Q}}}
& &  (\boldsymbol{W}{\boldsymbol{p}}_{\text{x}_{kI/Q}})^{\text{T}}(\boldsymbol{W}{\boldsymbol{p}}_{\text{x}_{kI/Q}})\\
& \text{subject to:}
& & \boldsymbol{B}_{k}\boldsymbol{p}_{\text{x}_{kI/Q}} \preceq  -\gamma(\text{SER}) \boldsymbol{a} \text{.}
\end{aligned}
\end{equation}
Considering the Gray coding for TI ZX modulation from \cite{Viveros_2024}. The approximately upper bound probability of error can be presented in terms of the bit error rate (BER) such that
{
\begin{align}
\label{BER_ub}
    \text{BER}_{\text{ub}} \simeq 
 \frac{\text{SER}_{\text{ub}}}{n_{s}},
\end{align}}
where $n_{s}$ corresponds to the number of bits per transmit symbol. For $M_{\text{Rx}}=3$, $2$ bits are mapped in one symbol, and for $M_{\text{Rx}}=2$, $3$ bits are mapped in 2 symbols.

Finally, in Algorithm~\ref{alg:PGD}, the process to obtain the temporal precoding vector ${\boldsymbol{p}}_{\text{x}_k}$ is summarized.
\begin{algorithm}[t]
		\caption{{Proposed QoS-based precoding algorithm to obtain ${\boldsymbol{p}}_{\text{x}_k}$}}
		\label{alg:PGD}
				\begin{algorithmic}[1]
                \STATE Define $M_\mathrm{Rx}$\\
                \STATE Construct $\boldsymbol{c}_{\text{out}}$\\
                \STATE Establish target SER according to Table~\ref{tab:gammaSERMRX3} or Table~\ref{tab:gammaSERMRX2}\\
                \STATE Solve Equation \eqref{eq:convex1b}\\
                 \STATE Obtain temporal precoding vector ${\boldsymbol{p}}_{\text{x}_k}$
                 \\                  
		\end{algorithmic}
		\end{algorithm}

%% file: numerical_Results.tex
\section{Simulation Results}
\label{sec:num_results}

In this section, we carry out numerical simulations to illustrate the performance of the proposed techniques. In particular, a comparison between the semi-analytical and numerical SER for the QOS precoding method is presented in this section. Alternatively, channel codes such as low-density parity-check codes \cite{ldpc,bfpeg,memd} could also be considered. The transmit filter $g_{\text{Tx}}(t)$ is an RC filter and the receive  filter $g_{\text{Rx}}(t)$  is an RRC filter, where the roll-off factors are $\epsilon_{\text{Tx}} = \epsilon_{\text{Rx}} = 0.22$. The bandwidth is defined with $W_{\text{Rx}}=W_{\text{Tx}} =\left ( 1 + \epsilon_{\text{Tx}} \right )/T$. The channel matrix $\boldsymbol{H}$ is complex Gaussian distributed with unit variance. Furthermore, different scenarios are considered with respect to $M_{\text{Rx}}$, $N_\text{t}$ and $N$. It is considered that $\gamma$ and the noise spectral density $N_{0}$, are related by $\gamma = c N_{0}$ where $c$ is a positive scalar value and $N_{0}=\sigma_{\text{n}}^2$. The required SNR, is defined as
{\begin{align}
\label{SNR}
    \text{SNR}_{\text{Req}} = 
 \frac{E_{\text{Tx}}}{N_{\text{q}}N_{0}\left ( 1 + \epsilon_{\text{Tx}} \right )}.
\end{align}}

\begin{figure}[H]
\begin{center}
\includegraphics[height=8cm, width=10cm]{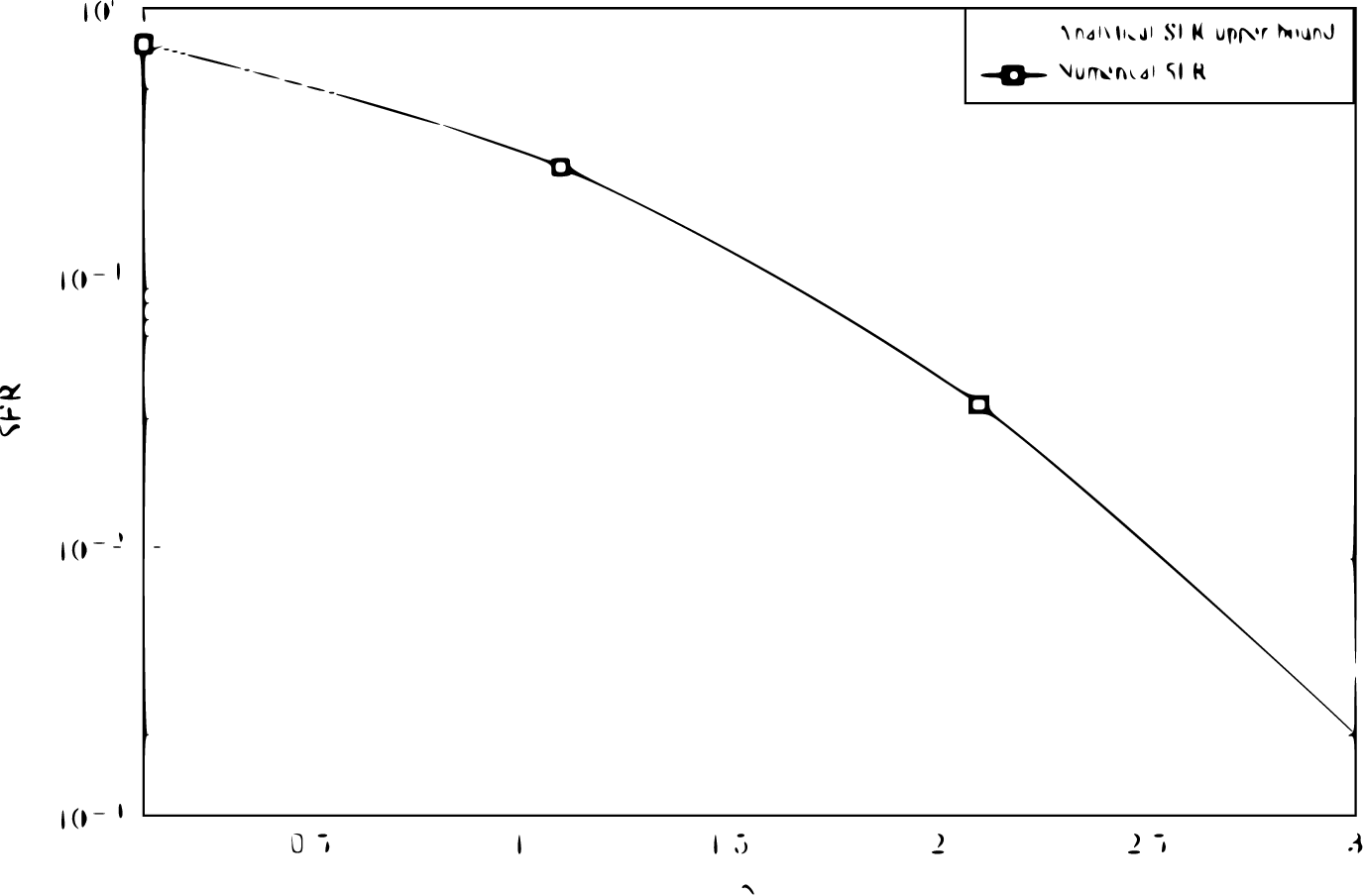}
\caption{Semi-analytical and numerical SER comparison for the MMDDT precoding method with $M_{\text{Rx}}= M_{\text{Tx}}= 3$, $N=1$ and $\sigma_{n}^2 = 1$.} 
\label{fig:analitical_QP3}    
\end{center}
\end{figure}

\begin{figure}[H]
\begin{center}
\includegraphics[height=8cm, width=10cm]{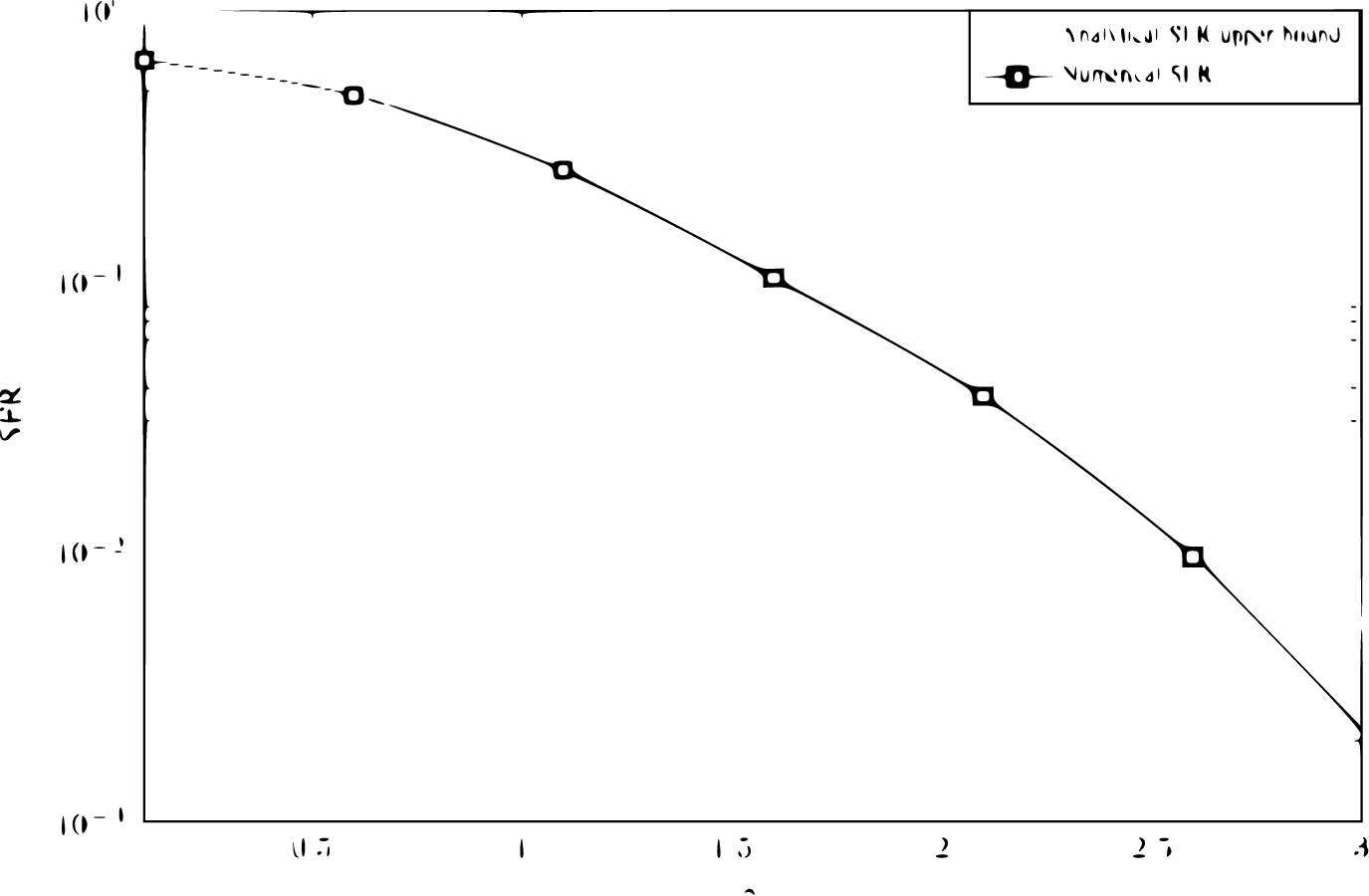}
\caption{Semi-analytical and numerical SER comparison for the MMDDT precoding method with $M_{\text{Rx}}= M_{\text{Tx}}= 2$, $N=2$ and $\sigma_{n}^2 = 1$.} 
\label{fig:analitical_QP2}    
\end{center}
\end{figure}

Fig.~\ref{fig:analitical_QP3} and Fig.~\ref{fig:analitical_QP2} compare the numerical SER for the QOS precoding with \eqref{eq:convex1} and the semi-analytical SER upper bound with noise variance $\sigma_{n}^2 = 1$ for $M_{\text{Rx}}=3$ and $M_{\text{Rx}}=2$, respectively. Due to the Gray coding for TI ZX modulation,  $1$ symbol for $M_{\text{Rx}}=3$ and sequences of $2$ symbols for $M_{\text{Rx}}=2$ were considered.

The SER for numerical and semi-analytical results considers the same value of $\gamma$. However, for the semi-analytical method, all the codewords are assumed to be constructed with $\gamma$. In the case of the QOS precoding method, $\gamma$ corresponds to the minimum distance to the decision threshold so samples can be larger than $\gamma$. 

\begin{figure}[H]
\begin{center}
\includegraphics[height=8cm, width=10cm]{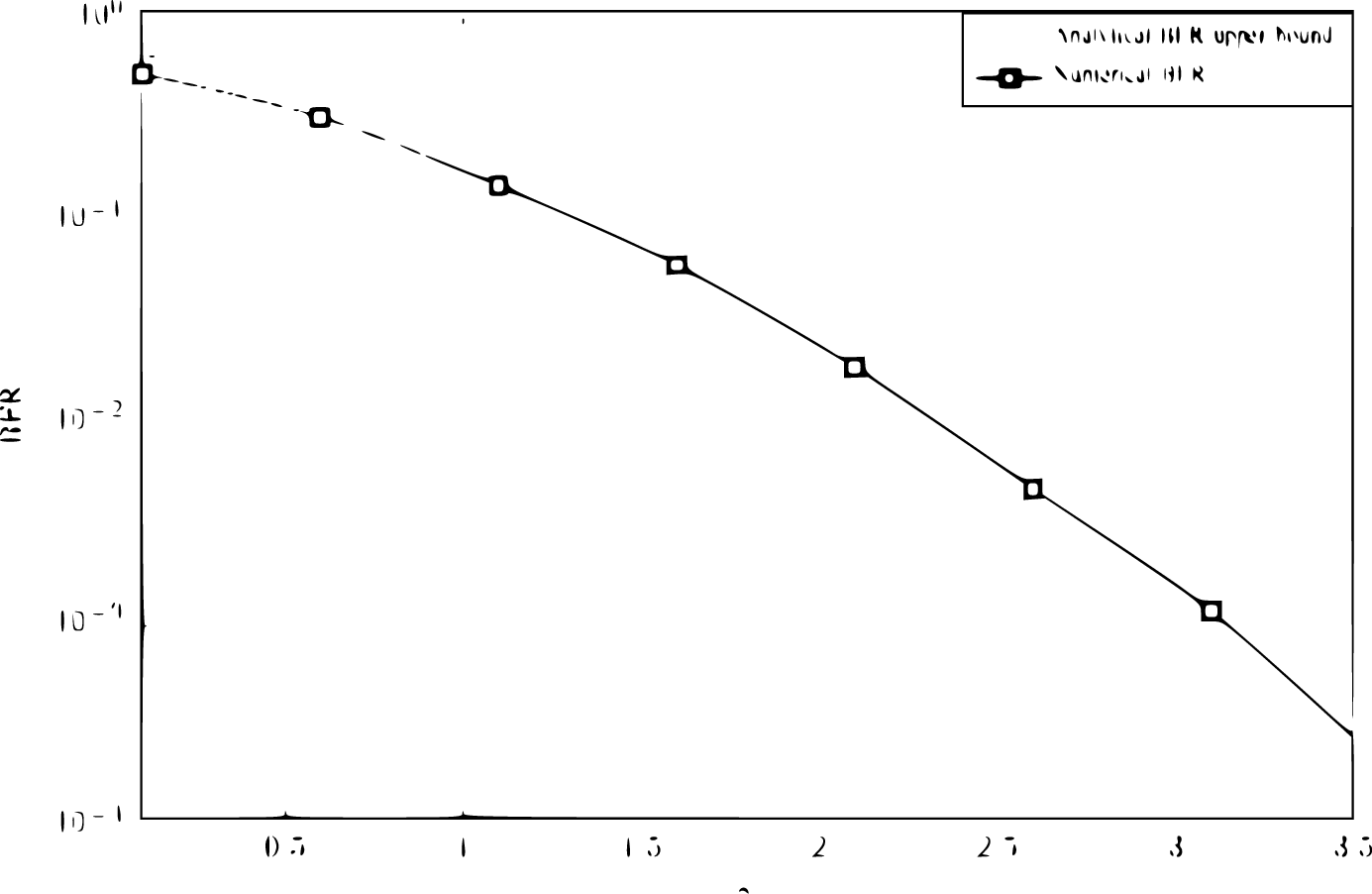}
\caption{Semi-analytical and numerical BER comparison for the MMDDT precoding method with $M_{\text{Rx}}= M_{\text{Tx}}= 3$, $N=1$ and $\sigma_{n}^2 = 1$.}
\label{fig:analitical_BER_QP3}    
\end{center}
\end{figure}

\begin{figure}[H]
\begin{center}
\includegraphics[height=8cm, width=10cm]{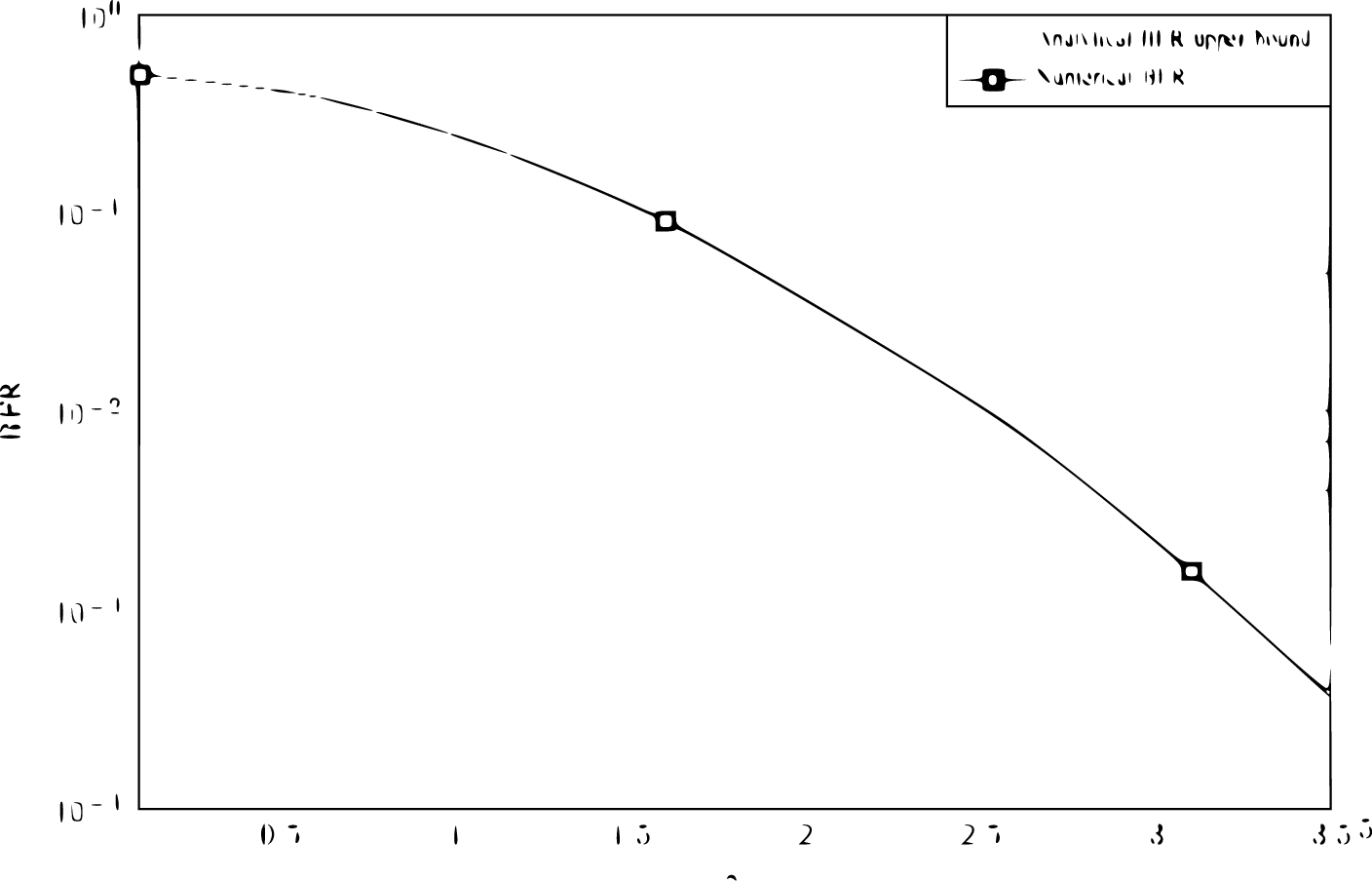}
\caption{Semi-analytical and numerical BER comparison for the MMDDT precoding method with $M_{\text{Rx}}= M_{\text{Tx}}= 2$, $N=2$ and $\sigma_{n}^2 = 1$.}
\label{fig:analitical_BER_QP2}    
\end{center}
\end{figure}

{Moreover, considering \eqref{BER_ub}, the results from Fig.~\ref{fig:analitical_QP3} and Fig.~\ref{fig:analitical_QP2} are also presented in terms of the BER in Fig.~\ref{fig:analitical_BER_QP3} and Fig.~\ref{fig:analitical_BER_QP2} for $M_{\text{Rx}}=3$ and $M_{\text{Rx}}=2$, respectively. Note that the analytical upper bound BER corresponds to an approximation of the real upper bound since the Gray coding for TI ZX could imply a negligible loss of information.}

\begin{figure}[H]
\begin{center}
\includegraphics[height=8cm, width=10cm]{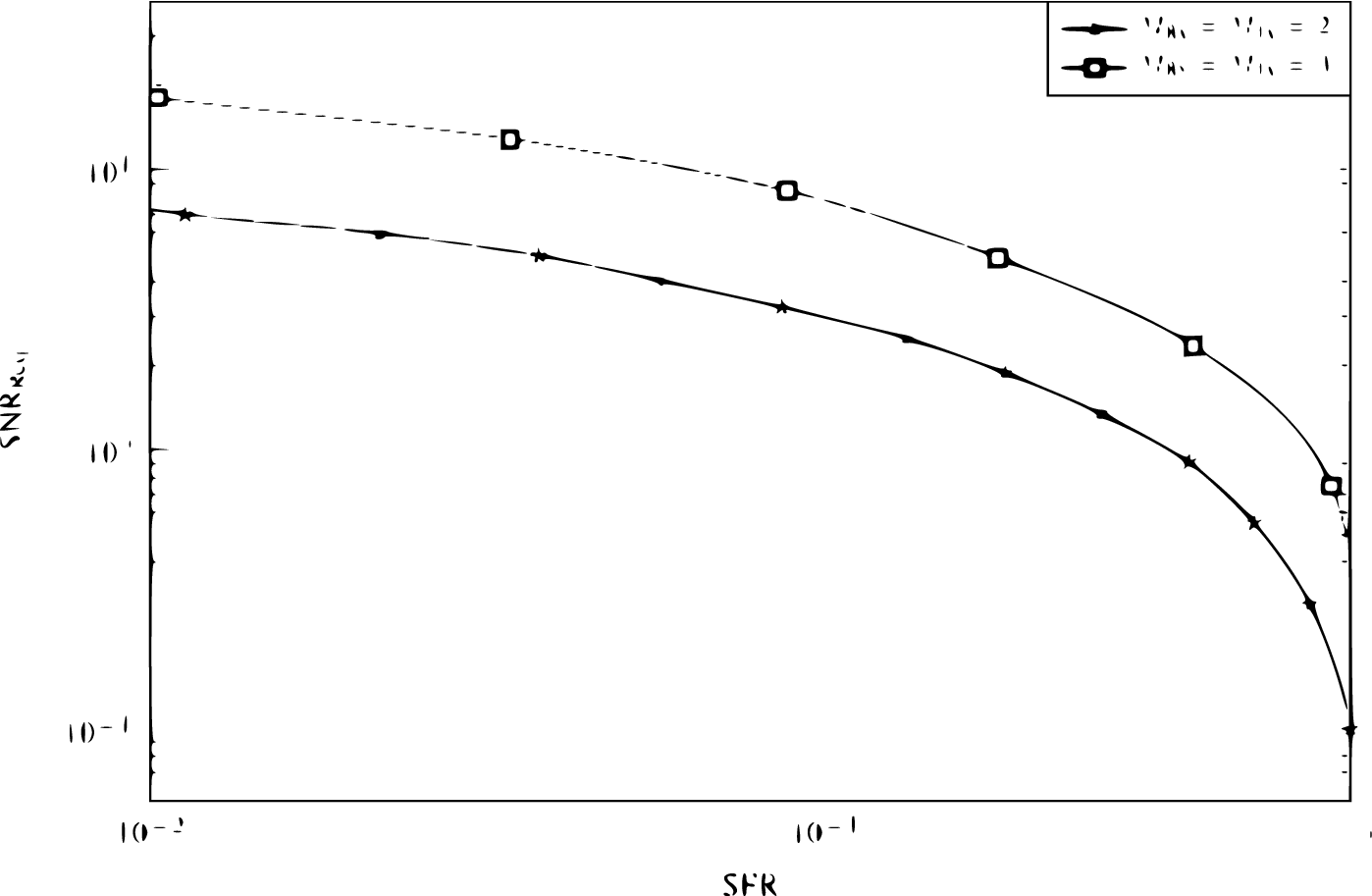}
\caption{{$\text{SNR}_{\text{Req}}$ vs. SER for
$N=2$, $\sigma_{n}^2 = 1$, $N_{\text{t}}= N_{\text{u}}=1$.}} 
\label{fig:powerSER2}    
\end{center}
\end{figure}

{Departing from the semi-analytical SER results in Fig.~\ref{fig:analitical_QP3} and Fig.~\ref{fig:analitical_QP2} the constraint $\gamma(\text{SER})$ considered in the optimization problem \eqref{eq:convex1b}, is shown in Table~\ref{tab:gammaSERMRX3} for $M_{\text{Rx}}= 3$ and Table~\ref{tab:gammaSERMRX2} for $M_{\text{Rx}}= 2$, considering $\sigma_{n}^{2} =1$. Fig.~\ref{fig:powerSER2} depicts the normalized transmit power $P_{\text{Tx}}$ \eqref{SNR}, where the total transmit energy $E_{\textrm{Tx}}$ is calculated with \eqref{Etx}. From Fig.~\ref{fig:powerSER2}, it can be seen that to achieve a lower SER value, higher transmitted power is required}. Moreover, to reach the same SER the set for $M_{\text{Rx}}= 3$ requires more energy than for $M_{\text{Rx}}= 2$.
\begin{table}[H]
\caption{$\gamma(\text{SER})$ for $M_{\text{Rx}}= 3$.} 
\label{tab:gammaSERMRX3}  
\begin{center}
\begin{tabular}{|cc|}
\hline
\multicolumn{2}{|c|}{$\gamma(\text{SER})$} \\ \hline
\multicolumn{1}{|c|}{SER}       & $\gamma$ \\ \hline
\multicolumn{1}{|c|}{$10^{-1}$}      & 1.75      \\ \hline
\multicolumn{1}{|c|}{$10^{-2}$}      & 2.65      \\ \hline
\multicolumn{1}{|c|}{$10^{-3}$}      & 3.35        \\ \hline
\multicolumn{1}{|c|}{$10^{-4}$}      & 3.9     \\ \hline
\multicolumn{1}{|c|}{$10^{-5}$}     & 4.45     \\ \hline
\multicolumn{1}{|c|}{$10^{-6}$}    & 4.8         \\ \hline
\end{tabular}
\end{center}
\end{table}
\begin{table}[H]
\caption{$\gamma(\text{SER})$ for $M_{\text{Rx}}= 2$.} 
\label{tab:gammaSERMRX2}  
\begin{center}
\begin{tabular}{|cc|}
\hline
\multicolumn{2}{|c|}{$\gamma(\text{SER})$} \\ \hline
\multicolumn{1}{|c|}{SER}       & $\gamma$ \\ \hline
\multicolumn{1}{|c|}{$10^{-1}$}       & 1.9      \\ \hline
\multicolumn{1}{|c|}{$10^{-2}$}       & 2.75      \\ \hline
\multicolumn{1}{|c|}{$10^{-3}$}       & 3.45     \\ \hline
\multicolumn{1}{|c|}{$10^{-4}$}       & 4     \\ \hline
\multicolumn{1}{|c|}{$10^{-5}$}      & 4.45       \\ \hline
\multicolumn{1}{|c|}{$10^{-6}$}     & 4.9          \\ \hline
\end{tabular}
\end{center}
\end{table}

\begin{figure}[H]
\begin{center}
\includegraphics[height=8cm, width=10cm]{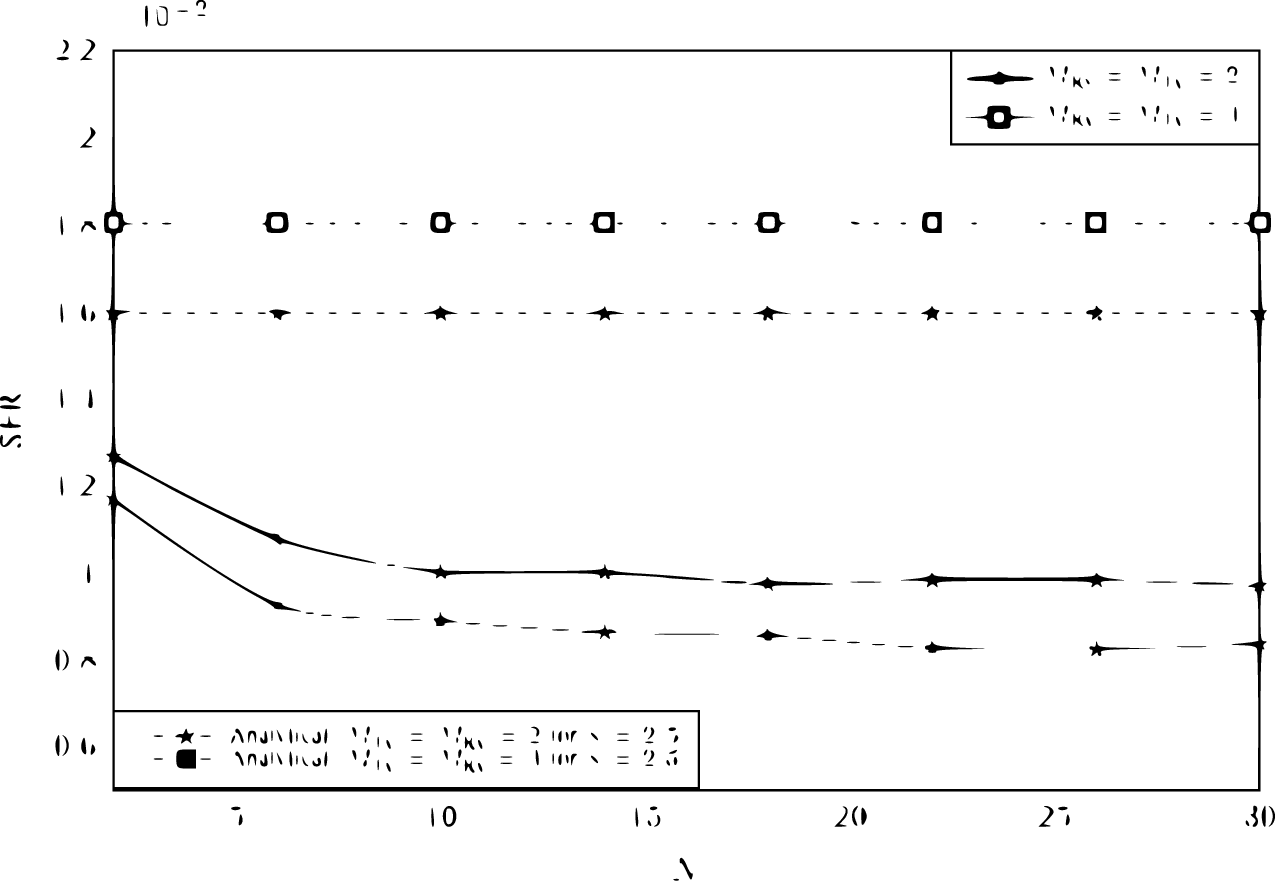}
\caption{{SER vs. $N$ for 
$\sigma^2 = 1$, $N_{\text{t}}= N_{\text{u}}=1$. The analytical curves are presented for $\gamma = 2.5$.}} 
\label{fig:SER_N}    
\end{center}
\end{figure}

\begin{figure}[H]
\begin{center}
\includegraphics[height=8cm, width=10cm]{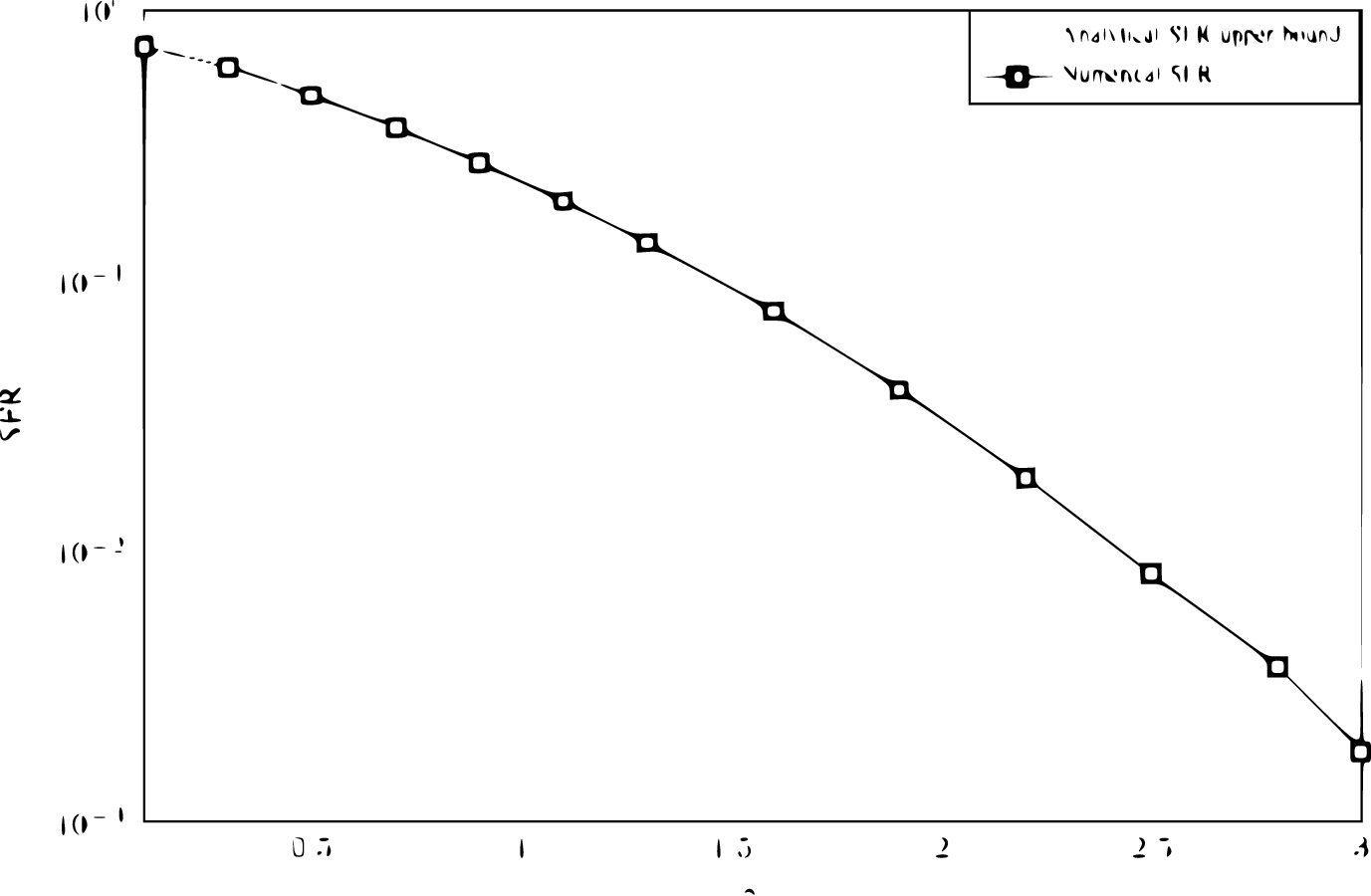}
\caption{Semi-analytical and numerical SER comparison for the MMDDT precoding method with $M_{\text{Rx}}= M_{\text{Tx}}= 3$, $N=20$, $\sigma_{n}^2 = 1$, $N_{\text{t}}=2$ and  $N_{\text{u}}=2$.} 
\label{fig:analitical_QP3_MU}    
\end{center}
\end{figure}

\begin{figure}[H]
\begin{center}
\includegraphics[height=8cm, width=10cm]{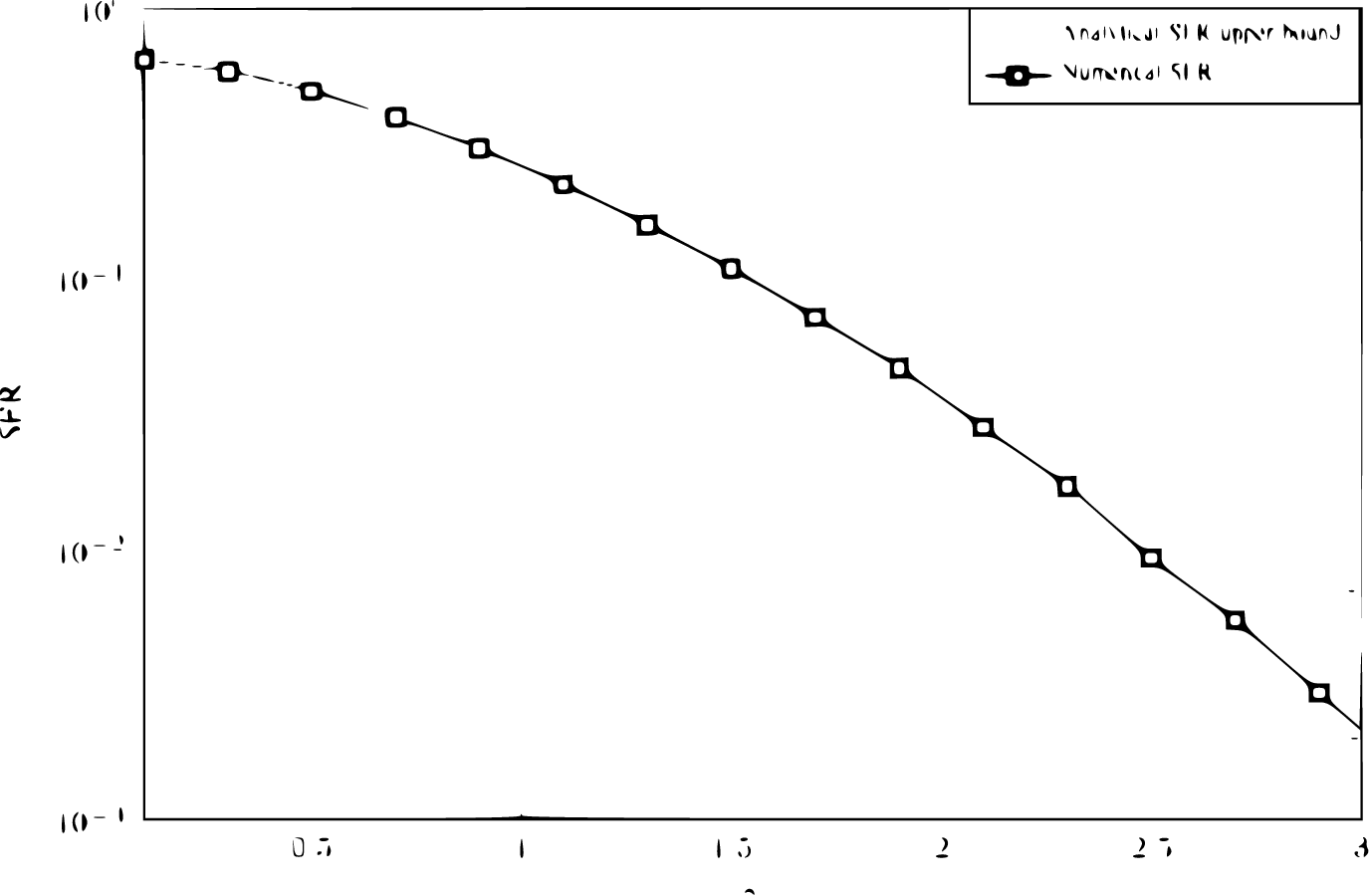}
\caption{Semi-analytical and numerical SER comparison for the MMDDT precoding method with $M_{\text{Rx}}= M_{\text{Tx}}= 2$, 
$N=20$, $\sigma_{n}^2 = 1$, $N_{\text{t}}=2$ and  $N_{\text{u}}=2$.} 
\label{fig:analitical_QP2_MU}    
\end{center}
\end{figure}

{In addition, Fig.~\ref{fig:SER_N}, shows the SER performance against the number of transmit symbols $N$ considering $M_{\text{Rx}}= M_{\text{Tx}}= 2$, $M_{\text{Rx}}= M_{\text{Tx}}= 3$,
$\sigma_{n}^2 = 1$ and $N_{\text{t}}= N_{\text{u}}=1$. The results show that increasing $N$  may result in a SER decreasing.} 
Moreover, Fig.~\ref{fig:analitical_QP3_MU} and Fig.~\ref{fig:analitical_QP2_MU} evaluate a multiuser MIMO scenario and compare the numerical SER for the QOS precoding with \eqref{eq:convex1b} and the semi-analytical SER upper bound with noise variance $\sigma_{n}^2 = 1$ for $M_{\text{Rx}}=3$ and $M_{\text{Rx}}=2$, respectively. For numerical evaluation, sequences of $N=20$ symbols,  $N_{\text{t}}=2$ and  $N_{\text{u}}=2$ were considered.

\begin{figure}[H]
\begin{center}
\includegraphics[height=8cm, width=10cm]{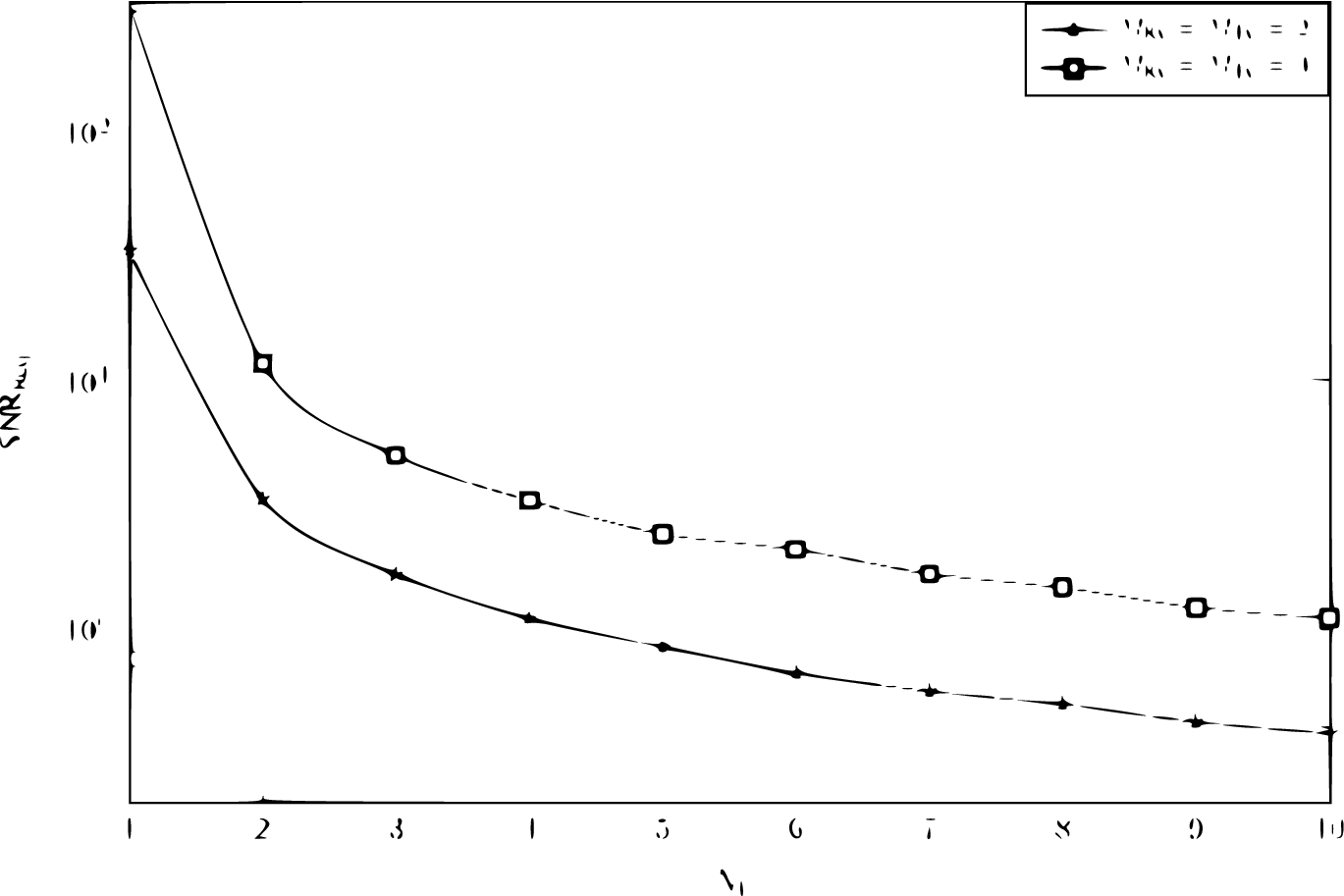}
\caption{$\text{SNR}_{\text{Req}}$ vs. 
 $N_{\text{t}}$ for $M_{\text{Rx}}= M_{\text{Tx}}= 2$ and $M_{\text{Rx}}= M_{\text{Tx}}= 3$ with
$N=10$, $\sigma_{n}^2 = 1$, and  $N_{\text{u}}=1$.} 
\label{fig:energy_ETX2}    
\end{center}
\end{figure}

Given the beamforming gain, in Fig.~\ref{fig:analitical_QP3_MU} and Fig.~\ref{fig:analitical_QP2_MU} the gap between the semi-analytical and numerical curves is larger than in  Fig.~\ref{fig:analitical_QP3} and Fig.~\ref{fig:analitical_QP2}. Furthermore, numerical results are presented in Fig.~\ref{fig:energy_ETX2} in terms of the transmit energy $E_{\text{Tx}_{\text{n}}}$ considering a different number of transmit antennas $N_{\text{t}}$.

\begin{figure}[H]
\begin{center}
\includegraphics[height=8cm, width=10cm]{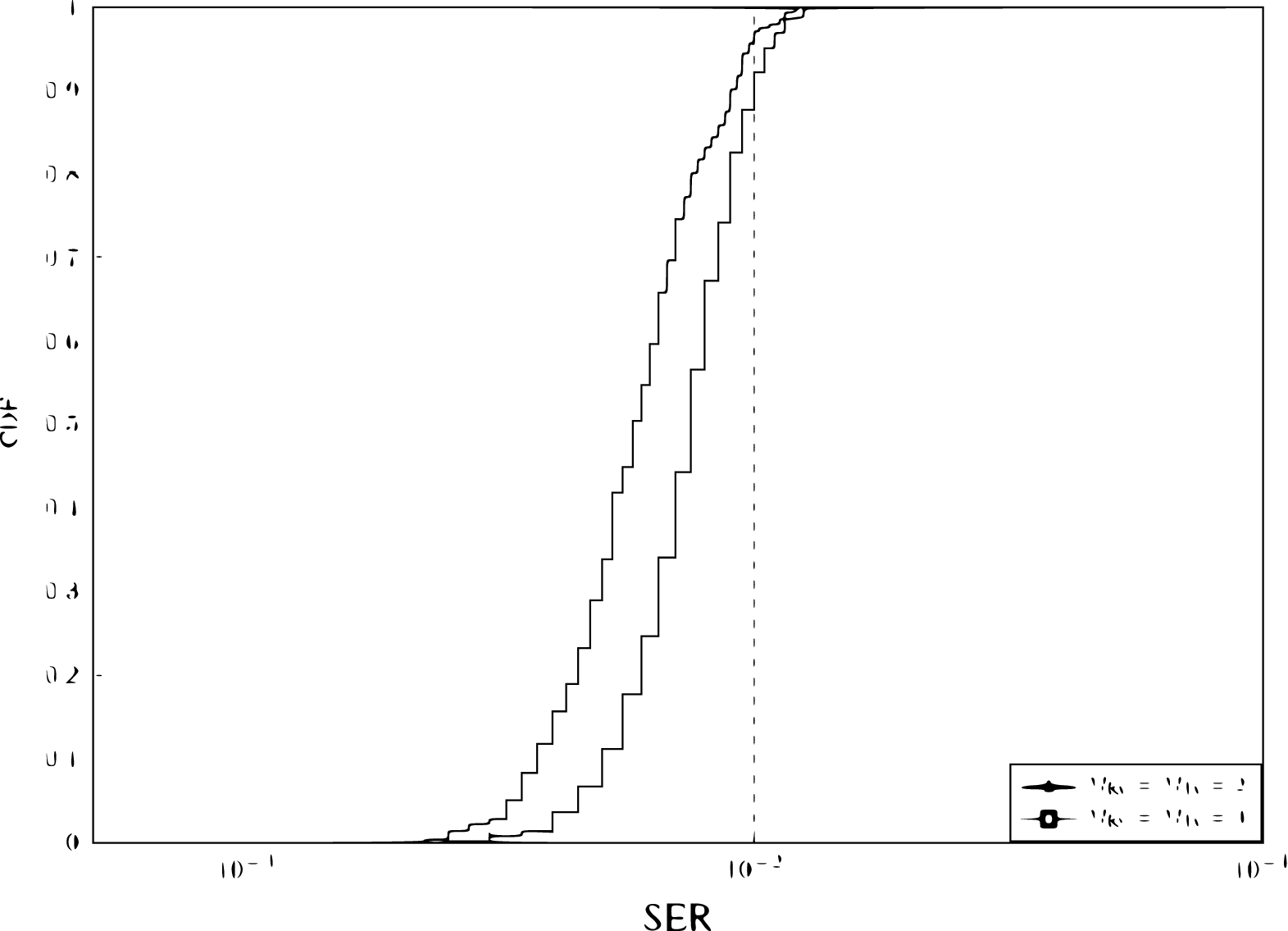}
\caption{CDF with $N=1$ for $M_{\text{Rx}}= M_{\text{Tx}}= 3$ and $N=2$ for $M_{\text{Rx}}= M_{\text{Tx}}= 2$, and $N_{\text{t}}=N_{\text{u}}=1$.} 
\label{fig:CDF}    
\end{center}
\end{figure}

\begin{figure}[H]
\begin{center}
\includegraphics[height=8cm, width=10cm]{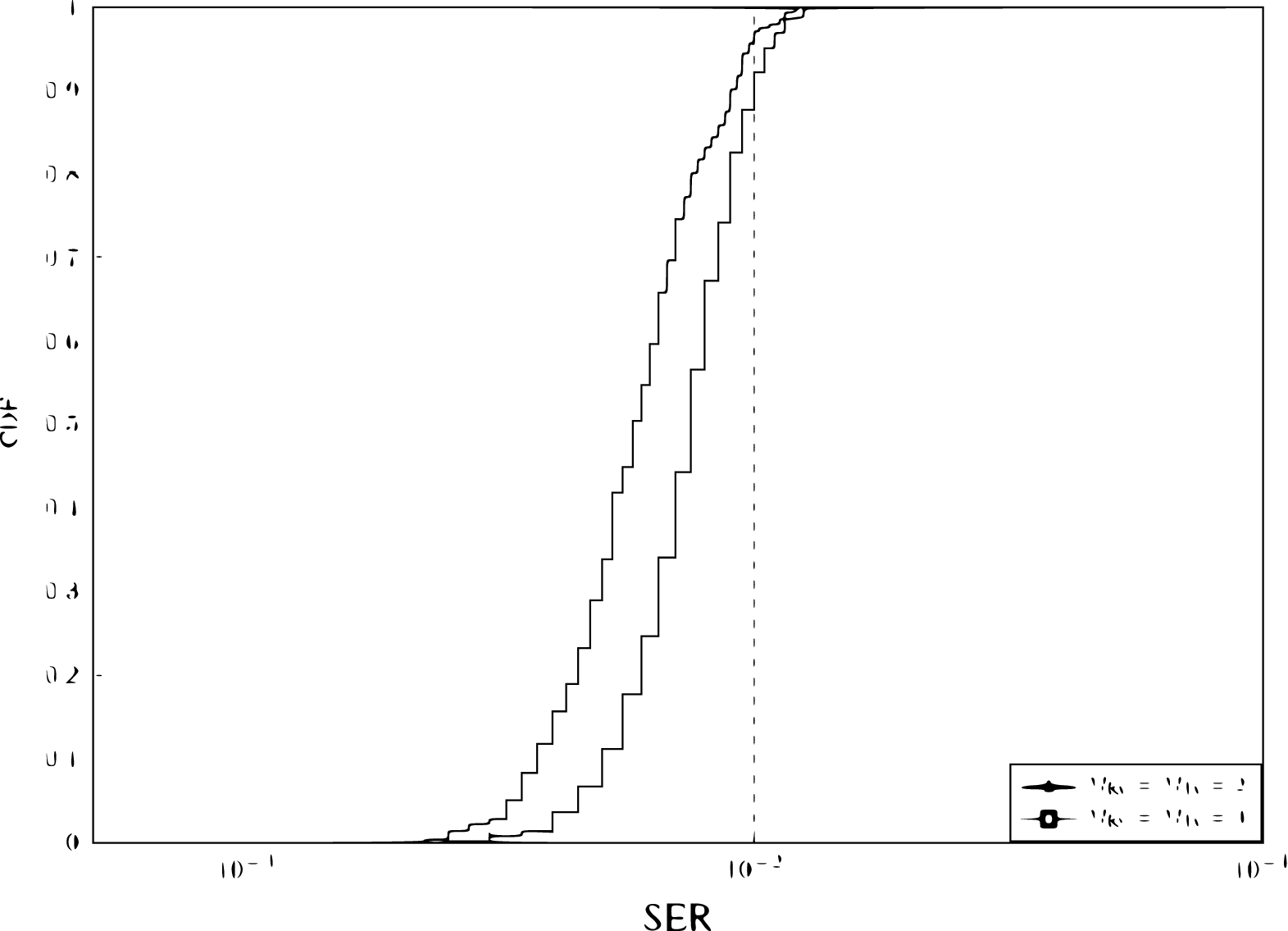}
\caption{CDF with $N=20$  and $N_{\text{t}}=N_{\text{u}}=2$} 
\label{fig:CDFMU}    
\end{center}
\end{figure}

From Fig.~\ref{fig:energy_ETX2}, it is possible to observe that transmitting $10$ symbols with $M_{\text{Rx}}= M_{\text{Tx}}= 3$ requires more power than transmitting the same number of symbols with $M_{\text{Rx}}= M_{\text{Tx}}= 2$ regardless of the number of transmit antennas. Moreover, for both cases presented, the power $P_{\textrm{Tx}}$ decreases when the number of transmit antennas increases. 
Finally, Fig.~\ref{fig:CDF} and Fig.~\ref{fig:CDFMU} depict the cumulative distribution function (CDF) of the SER, where the SER constraint in \eqref{eq:convex1b} is $10e-2$ for a single antenna user case and multiuser case, respectively.  Particularly, in both figures, the CDF shows a high probability of fulfilling the SER constraint for both studied scenarios. The parameters for results presented in Fig.~\ref{fig:CDF} are $N=1$ for $M_{\text{Rx}}= M_{\text{Tx}}= 3$ and $N=2$ for $M_{\text{Rx}}= M_{\text{Tx}}= 2$, both scenarios are simulated with $N_{\text{t}}=N_{\text{u}}=1$. The parameters for results presented in Fig.~\ref{fig:CDFMU} are $N=20$, $N_{\text{t}}=2$ and $N_{\text{u}}=2$.

%% file: conclusions.tex
\section{Conclusions}
\label{sec:conclusiones}
This work presented semi-analytical results regarding the SER for a precoding method based on a QOS constraint with time instance zero-crossing modulation. The considered system is a bandlimited multiuser MIMO downlink scenario, with 1-bit quantization and oversampling at the receiver. The QOS precoding method minimizes the transmitted energy while taking into account the quality of service constraint which is given in terms of a targeted SER.  The semi-analytical SER corresponds to an upper bound since it is considered that all the transmit samples have the same distance to the decision threshold. The probability of error has been calculated numerically through the cumulative distribution function of a multivariate normal distribution. 
The comparison between the numerical and semi-analytical results for $M_\mathrm{Rx} =2$ and $M_\mathrm{Rx} =3$ has shown that the semi-analytical method corresponds to an upper bound, which is consistent with established theory. The semi-analytical approach is suitable for precoding design and numerical simulation confirms that the proposed algorithms can attain the target SER.
